\documentclass[a4paper]{article}

\def\withpics{1}

\usepackage{theorem}
\usepackage{amsmath}
\usepackage[psamsfonts]{amssymb}
\usepackage{latexsym}
\usepackage{cmmib57}
\if1\withpics
\usepackage{graphicx}
\usepackage{pstricks}
\usepackage{pst-coil}
\usepackage{pst-node}
\usepackage{float}
\usepackage{wrapfig}
\usepackage{epsf}

\newgray{meascolor}{.5}
\newgray{opercolor}{.5}
\newgray{entacolor}{.5}
\newgray{qbitcolor}{.1}
\newgray{mklight}{.7}
\newgray{mkfill}{.9}
\psset{unit=\unitlength}
\psset{arrowsize=2pt 5}
\psset{coilarm=.4}
\fi

\newcommand{\scr}[1]{\mathcal{#1}}
\def\idty{{\leavevmode{\rm 1\ifmmode\mkern -4.8mu\else\kern -.3em\fi
      I}}}
\renewcommand{\Bbb}[1]{\if1#1\idty\else\mathbb{#1}\fi}

\newcommand{\kb}[1]{|#1\rangle\langle#1|}
\newcommand{\KB}[2]{|#1\rangle\langle#2|}
\newcommand{\ket}[1]{|#1\rangle}
\newcommand{\tr}{\operatorname{tr}}
\newcommand{\Id}{\operatorname{Id}}

\newcommand{\cb}{{\rm cb}}

\let\id\Id
\renewcommand{\epsilon}{\varepsilon}

\def\Cx{{\mathbb C}}\def\Nl{{\mathbb N}}

\def\HH{{\scr H}}\def\B{{\scr B}}

\def\Zd{{{\mathbb Z}_d}}
\def\cbnorm#1{\|#1\|_{\rm cb}}
\def\norm#1{\|#1\|}
\def\QC{Q} %quantum capacity

\newtheorem{thm}{Theorem}[section]
\newtheorem{defi}[thm]{Definition}
\newtheorem{prop}[thm]{Proposition}
\newtheorem{lem}[thm]{Lemma}
\newtheorem{kor}[thm]{Corollary}
\newenvironment{proof}{\par\noindent\textit{Proof.\ }}{\hfill $\Box$ \vspace{1em}}

\begin{document}

\title{How to correct small quantum errors}
\author{ M. Keyl\thanks{Electronic Mail: \tt{m.keyl@tu-bs.de}}
 {{}\ and\ } R.~F. Werner\thanks{Electronic Mail: \tt{r.werner@tu-bs.de}}
   \\[1ex]
  {\small Institut f{\"u}r Mathematische Physik, TU Braunschweig,}\\
  {\small Mendelssohnstr.3, 38106 Braunschweig, Germany.}}
\date{\today}
\maketitle

\begin{abstract}
  The theory of quantum error correction is a cornerstone of quantum information processing. It 
  shows that quantum data can be protected against decoherence effects, which otherwise would
  render many of the new quantum applications practically impossible. In this paper we give a self
  contained introduction to this theory and to the closely related concept of quantum channel capacities. We
  show, in particular, that it is possible (using appropriate error correcting schemes) to send a
  non-vanishing amount of quantum data undisturbed (in a certain asymptotic sense) through a noisy
  quantum channel $T$, provided the errors produced by $T$ are small enough.
\end{abstract}

\vspace{4em}

\begin{center}
  \begin{minipage}{12cm}
    This text is part of a volume entitled: ``\emph{Coherent evolution in noisy environments}'' to be
    published in \emph{Lecture notes in physics},  Springer Verlag,
    \texttt{http://link.springer.de/series/lnpp/}, Copyright: Springer Verlag, Berlin, Heidelberg, New
    York 
  \end{minipage}
\end{center}

\clearpage

% $Log: DDQCapIntro.tex,v $
% Revision 1.5  2002/05/03 12:04:41  michael
% Final version.
%
% Revision 1.4  2002/04/11 10:15:04  michael
% First complete draft. Several improvements are still necessary.
%
% Revision 1.3  2002/03/26 17:56:40  michael
% Many, many, many changes and additons by Reinhard.
%
% Revision 1.2  2002/03/06 15:30:12  michael
% Some references added. Knill and Laflamme stuff now complete.
%
% Revision 1.1  2002/03/05 11:03:18  michael
% Divided the project into several files. Intoduction added.
%
%
% $Id: DDQCapIntro.tex,v 1.5 2002/05/03 12:04:41 michael Exp $

%%%%%%%%%%%%%%%%%%%%%%%%%%%%%%%%%%%%%%%%%%%%%%%%%%%%%%%%%%%%%%%%%%%%%%%%%%%%%%%%%%%%%%%%%%%%%%%%%%%%
\section{Introduction}
%%%%%%%%%%%%%%%%%%%%%%%%%%%%%%%%%%%%%%%%%%%%%%%%%%%%%%%%%%%%%%%%%%%%%%%%%%%%%%%%%%%%%%%%%%%%%%%%%%%%

Controling decoherence is one of the key problems for making
quantum information processing and quantum computation work. From
the outset, when Peter Shor announced his algorithm
\cite{Shor94,Shor97}, many physicists felt that somewhere there
would be a price to pay for the miraculous exponential speedup.
For example, if the algorithm would require exponentially good adherence to specifications for  
the quantum circuitry and exponentially low noise levels, it would have been totally useless. Indeed
it is far from easy to show that it does not make such
requirements.

In this article we look at the simpler, but equally fundamental
problem of quantum information transmission or storage. Is it
possible to encode the quantum data in such a way that even after
some degradation they can be restored nearly perfectly by a
suitable decoding operation? Assuming that the degrading
decoherence effects are small to begin with, can restoration be
made nearly perfect?

For classical information it is very simple to do this, namely by
redundant coding. If we want to send one bit through a noisy
channel, we can reduce errors by sending it three times and
deciding by majority vote which value we take at the output.
Clearly, if errors have a small probability $\varepsilon$ for a
single channel, they will have order $\varepsilon^2$ for the
triple channel, because we go wrong only when two independent
errors occur. Unfortunately, such a scheme cannot work in the
quantum case because it involves a copying operation, which is
forbidden by the No-Cloning Theorem \cite{WooZu}. So we have to look
for subtler ways of distributing quantum information among several
systems and thereby reducing the probability of errors. Indeed
such schemes exist \cite{CSQECC,StQECC} and are the subject of the exciting new
field of quantum error correcting codes.

The efficiency of such a scheme is measured by two parameters,
namely how many uses of the noisy channel are required, and the
error level after correction. The above simple classical scheme
can be iterated to get the errors for a single bit down to
$\varepsilon^{2^n}$ with $3^n$ parallel uses of the channel. This
is a large overhead to correct a single bit. Better procedures
work classically by coding several bits at a time, and one can
manage to make errors as small as desired with only a finite
overhead per bit. The minimal required overhead (or rather its
inverse) is, in fact, the central quantity of the coding theory
\cite{Shannon48} for noisy channels: one defines the {\it
capacity} of a channel as the number of bit transmissions per use
of the channel, in an optimal coding scheme for messages of length
$L\to\infty$ with the property that the error probability goes to
zero in this limit.

It is not a priori clear that the notion of channel capacity makes
sense for quantum information, i.e. that the capacity of a channel which produces only small errors is
nonzero and close to that of the ideal (errorless) channel. This is indeed not even evident from most
existing presentations of the theory of quantum error correcting codes. Papers which address this problem
at least for special cases like depolarizing channels are \cite{DiVSS,GottD} and \cite[Sec 7.16.2]{P219}
while the general case is treated more recently in \cite{Hamada01,MaUy01}. The 
purpose of this paper is less the presentation of new results but to show in an elementary and
self-contained way that small quantum errors can be corrected with an asymptotically small effort. 
To this end the paper is organized as follows. We first review the basic
notions concerning quantum channels (Section \ref{sec:quantum-channels}), and give an abstract definition
of the capacity together with some elementary properties (Section \ref{sec:channel-capacities}). Then we
discuss the theory of error correcting codes (Section \ref{sec:quant-error-corr}) and a particular scheme
to construct such codes which is based on graph theory (Section \ref{sec:graph-codes}). In Section
\ref{sec:discr-cont-error} and \ref{sec:coding-random-graphs} we apply this scheme to channel capacities
and finally we draw our conclusions in Section \ref{sec:conclusions}.

% $Log: DDQCapChan.tex,v $
% Revision 1.4  2002/04/11 10:15:04  michael
% First complete draft. Several improvements are still necessary.
%
% Revision 1.3  2002/03/26 17:56:40  michael
% Many, many, many changes and additons by Reinhard.
%
% Revision 1.2  2002/03/05 19:00:12  michael
% Shortened the "channel" part and started with new Knill and Laflamme stuff.
%
% Revision 1.1  2002/03/05 11:03:18  michael
% Divided the project into several files. Intoduction added.
%
%
% $Id: DDQCapChan.tex,v 1.4 2002/04/11 10:15:04 michael Exp $

%%%%%%%%%%%%%%%%%%%%%%%%%%%%%%%%%%%%%%%%%%%%%%%%%%%%%%%%%%%%%%%%%%%%%%%%%%%%%%%%%%%%%%%%%%%%%%%%%%%%
\section{Quantum channels}
\label{sec:quantum-channels}
%%%%%%%%%%%%%%%%%%%%%%%%%%%%%%%%%%%%%%%%%%%%%%%%%%%%%%%%%%%%%%%%%%%%%%%%%%%%%%%%%%%%%%%%%%%%%%%%%%%%

According to the rules of quantum mechanics, every kind of quantum
systems is associated with a Hilbert space $\scr{H}$, which for
the purpose of this article we can take as finite dimensional.
Since even elementary particles require infinite dimensional
Hilbert spaces, this means that we are usually only trying to
coherently manipulate a small part of the system. The simplest
quantum system has a two dimensional Hilbert space
$\scr{H}=\Cx^2$, and is called a \emph{qubit}, for `quantum bit'.
The observables of the system are given by bounded operators. This
space will be denoted by $\scr{B}(\scr{H})$. The preparations
(states) are given by density operators
$\rho\in\scr{B}_*(\scr{H})$, where the latter denotes the space of
trace class operators on ${\scr H}$.  Of course, on finite
dimensional Hilbert spaces all linear operators are bounded {\it
and} trace class. So we use this notation mostly  to keep track of
the distinction between spaces of observables and spaces of
states.

A \emph{quantum channel}, which transforms input systems described
by a Hilbert space  $\scr{H}_1$ into output systems described by a
(possibly different) Hilbert space $\scr{H}_2$ is represented
mathematically by a \emph{completely  positive, unital map}
$T:\scr{B}(\scr{H}_2) \to \scr{B}(\scr{H}_1)$. Each $T$ can be
written in the form \cite{Kraus}
\begin{equation} \label{eq:2}
  T(A) = \sum_{j=1}^n F_j^* A F_j,
\end{equation}
where the $F_j$ are (bounded) operators $\scr{H}_2 \to \scr{H}_1$,
called  \emph{Kraus operators}. The equivalence of this form to
the condition of complete positivity is a simple consequence of
the Stinespring theorem \cite{StSpr}.

The physical interpretation of $T$ is the following. The
expectation value of an $A$ measurement ($A \in
\scr{B}(\scr{H}_2)$) at the output side of the channel, on a
system which is initially in the state $\rho \in
\scr{B}_*(\scr{H}_1)$ is given in terms of $T$ by $\tr[\rho
T(A)]$. Alternatively we can introduce the map $T_*:
\scr{B}_*(\scr{H}_1) \to \scr{B}_*(\scr{H}_2)$ which is
\emph{dual} to $T$, i.e. $\tr[T_*(\rho) A] = \tr[\rho T(A)]$. It
is uniquely determined by $T$ (and vice versa) and we can say that
$T_*$ represents the channel in the \emph{Schr{\"o}dinger picture},
while $T$ provides the \emph{Heisenberg picture} representation.

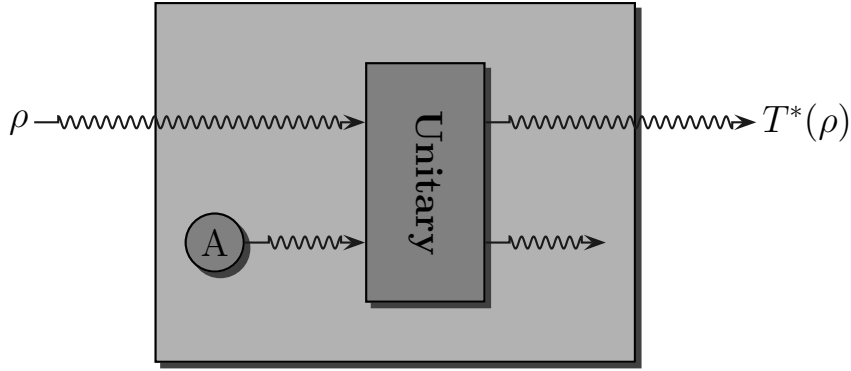
\begin{figure}[t]
  \begin{center}
    \begin{picture}(16,6)
      \rput[bl](-0.5,0){
      \psframe[fillcolor=mklight,fillstyle=solid,shadow=true](4,0)(12,6)
      \psframe[fillcolor=entacolor,fillstyle=solid,shadow=true](7.5,1)(9.5,5)
      \pccoil[linecolor=qbitcolor,coilaspect=0,coilheight=1,coilwidth=.2]{->}(2,4)(7.5,4)
      \pccoil[linecolor=qbitcolor,coilaspect=0,coilheight=1,coilwidth=.2]{->}(9.5,4)(14,4)
      \pscircle[fillcolor=meascolor,fillstyle=solid,shadow=true](5,2){0.5}
      \rput(5,2){\Large A}
      \pccoil[linecolor=qbitcolor,coilaspect=0,coilheight=1,coilwidth=.2]{->}(5.5,2)(7.5,2)
      \pccoil[linecolor=qbitcolor,coilaspect=0,coilheight=1,coilwidth=.2]{->}(9.5,2)(11.5,2)
      %\pccoil[coilheight=1,coilwidth=.2,coilarm=3pt]{*-*}(10.5,4)(10.5,2)
      \rput(8.5,3){\rotateright{\Large \bf Unitary}}
      \rput[r](1.9,4){\Large $\rho$}
      \rput[l](14.1,4){\Large $T^*(\rho)$}}
    \end{picture}
    \caption{Noisy channel}
    \label{fig:noisy-channel}
  \end{center}
\end{figure}

Let us consider now the special case that $\scr{H}_1 = \scr{H}_2 =
\scr{H}$. For example $T$ describes the transmission of photons
through an optical fiber or the storage in some sort of quantum
memory. Ideally we would prefer  channels which do not affect the
information at all, i.e. $T = \Id$, the identity map on
$\scr{B}(\scr{H})$. We will call this case the \emph{ideal
channel}. In real situations, however, interaction with the
environment, i.e. additional, unobservable degrees of freedom, can
not be avoided. The general structure of such a \emph{noisy
channel} is given by
\begin{equation} \label{eq:1}
  \rho \mapsto T_*(\rho) = \tr_\scr{K} \bigl(U (\rho \otimes \rho_0) U^*\bigr).
\end{equation}
where $U: \scr{H} \otimes \scr{K} \to \scr{H} \otimes \scr{K}$ is
a unitary operator describing the common evolution of the system
(Hilbert space $\scr{H}$) and the environment (Hilbert space
$\scr{K}$) and $\rho_0 \in \scr{S}(\scr{K})$ is the initial state
of the environment (cf. Figure \ref{fig:noisy-channel}). Note that
each $T$ can be represented in this way (this is again an easy
consequence of the Stinespring theorem), however there are in
general many possible choices for such an ``ancilla
representation''.

\if1\withpics
\fi

% $Log: DDQCapQC.tex,v $
% Revision 1.4  2002/05/03 12:04:41  michael
% Final version.
%
% Revision 1.3  2002/04/11 10:15:04  michael
% First complete draft. Several improvements are still necessary.
%
% Revision 1.2  2002/03/26 17:56:40  michael
% Many, many, many changes and additons by Reinhard.
%
% Revision 1.1  2002/03/05 11:03:18  michael
% Divided the project into several files. Intoduction added.
%
%
% $Id: DDQCapQC.tex,v 1.4 2002/05/03 12:04:41 michael Exp $

%%%%%%%%%%%%%%%%%%%%%%%%%%%%%%%%%%%%%%%%%%%%%%%%%%%%%%%%%%%%%%%%%%%%%%%%%%%%%%%%%%%%%%%%%%%%%%%%%%%%
\section{Channel capacities}
\label{sec:channel-capacities}
%%%%%%%%%%%%%%%%%%%%%%%%%%%%%%%%%%%%%%%%%%%%%%%%%%%%%%%%%%%%%%%%%%%%%%%%%%%%%%%%%%%%%%%%%%%%%%%%%%%%

As we have already pointed out in the introduction, the \emph{capacity} of a quantum channel is, roughly
speaking, the number of qubits transmitted per channel usage. In this section we will come to a more
precise description.

%%%%%%%%%%%%%%%%%%%%%%%%%%%%%%%%%%%%%%%%%%%%%%%%%%
\subsection{The cb-norm}
\label{sec:cb-norm}
%%%%%%%%%%%%%%%%%%%%%%%%%%%%%%%%%%%%%%%%%%%%%%%%%%

As a first step we need a measure for the difference between a noisy channel 
$T:\scr{B}(\scr{H})\to\scr{B}(\scr{H})$ and its ideal counterpart.
There are several mathematical ways of expressing this, which turn
out to be equivalent for our purpose. %\cite{footnote}. 
We find it
most convenient to take a certain norm difference, i.e., to
consider $\cbnorm{T-\id}$ as a quantitative description of the
noise level in $T$, where $\cbnorm\cdot$ denotes a certain norm,
called the norm of \emph{complete boundedness} (``cb-norm'' for
short). Its physical meaning is that of the largest difference
between probabilities measured in two experimental setups,
differing only by the substitution of $T$ by $\Id$. Since this
setup may involve further subsystems, and the measurement and
preparation may be entangled with the systems under consideration,
we have to take into account such additional systems in the
definition of the norm. For a general linear operator
$T:\scr{B}(\HH_2)\to\scr{B}(\HH_1)$ we set
\begin{equation}
  \cbnorm T = \sup\Bigl\{ \norm{(T\otimes\id_n)(A)}
                  \Bigm\vert n\in\Nl;
                             A\in\scr{B}(\HH_2\otimes\Cx^n);
                              \norm A\leq1
                  \Bigr\}\;.
\end{equation}
The cb-norm improves the sometimes annoying property of the usual
operator norm that quantities like $\|T \otimes
\Id_{\scr{B}(\Bbb{C}^d)}\|$ may increase with the dimension $d$.
On infinite dimensional Hilbert spaces $\cbnorm T$ can be infinite
although the supremum for every fixed $n$ is finite. A particular
example for a map with such a behavior is the transposition. A map
with finite cb-norm is therefore called completely bounded. In a
finite dimensional setup each linear map is completely bounded.
For the transposition $\Theta$ on $\Bbb{C}^d$ we have in
particular $\|\Theta\|_\cb = d$. The cb-norm has some nice
features which we will use frequently. This includes its
multiplicativity $\cbnorm{T_1 \otimes T_2} = \cbnorm{T_1}\;
\cbnorm{T_2}$ and the fact that  $\cbnorm{T} = 1$ for every
channel. For more properties of the cb-norm we refer to
\cite{Paulsen}.

%%%%%%%%%%%%%%%%%%%%%%%%%%%%%%%%%%%%%%%%%%%%%%%%%%
\subsection{Achievable rates and capacity}
%%%%%%%%%%%%%%%%%%%%%%%%%%%%%%%%%%%%%%%%%%%%%%%%%%

How can we reduce the error level $\cbnorm{T-\id}$? As an example,
consider a small unitary rotation, i.e., $T(X)=U^*XU$, with
$\cbnorm{T-\id}\leq2\norm{U-\idty}$ small. Then if we know $U$, it
is easy to correct $T$ by the inverse rotation, either before $T$,
as an ``encoding'', or afterwards, as a ``decoding'' operation.
More generally, we may use both, i.e., we are trying to make the
combination $ETD\approx\id$, by careful choice of the channels $E$
and $D$. Note that in this way we may look at channels $T$, which
have different input and output spaces, and hence cannot be
compared directly with the ideal channel on any system. For such
channels there is no intrinsic way of defining ``errors'' as
deviations from a desired standard. Moreover, we are free to
choose the Hilbert space $\HH_0$ such that
$ETD:\scr{B}(\HH_0)\to\scr{B}(\HH_0)$. For the product $ETD$ to be
defined, it is then necessary that
$D:\scr{B}(\HH_0)\to\scr{B}(\HH_2)$ and
$E:\scr{B}(\HH_1)\to\scr{B}(\HH_0)$. The best error level we can
achieve deserves its own notation. We define
\begin{equation}
  \Delta(T,M) = \inf_{E,D}\cbnorm{ETD - \Id}\;,
\end{equation}
where the infimum is taken over all encodings $E$ and decodings
$D$ and $M$ is the dimension of the space $\HH_0$. Now for longer
messages, e.g., a message of $m$ qubits (so that $M=2^m$) we need
to use the channel more often. In the language of classical
information theory, we are using longer code words, say of length
$n$. The error for coding $m$ qubits through $n$ uses of the
channel $T$ is then $\Delta(T^{\otimes n},2^m)$. Can we make this
small while retaining a good rate $m/n$ of bits per channel?
Clearly there will be a trade-off between rate and errors, which
is the basis of the following Definition. The notation $\lfloor x
\rfloor$, read ``floor $x$'', denotes the largest integer $\leq
x$.

\begin{defi} \label{def:1}
$c \geq 0$ is called \emph{achievable rate} for $T$, if
\begin{equation}
    \lim_{n \to \infty} \Delta(T^{\otimes n}, \lfloor 2^{ cn }\rfloor) = 0.
\end{equation}
  The supremum of all achievable rates is called the
  \emph{quantum-capacity} of $T$ and is denoted by $Q(T)$.
\end{defi}

Because $c=0$ is always an achievable rate we have $Q(T)\geq0$. On
the other hand, if every $c> 0$ is achievable we write
$Q(T)=\infty$.

Often a coding scheme construction does not work for arbitrary
integers, but only for specific values of $n$, or the dimension of
the coding space. However, this is no serious restriction, as the
following Lemma shows.

\begin{lem} \label{lem:1} Let $(n_\alpha)_{\alpha\in\Nl}$ be a strictly increasing
sequence of integers such that $\lim_\alpha
n_{\alpha+1}/n_\alpha=1$. Suppose $M_\alpha$ are integers such
that
 $\lim_\alpha\Delta(T^{\otimes n_\alpha},M_\alpha)=0$. Then
any
\begin{equation}
  c< \liminf_\alpha \frac{\log_2M_\alpha}{n_\alpha}
\end{equation}
is an admissible rate.  Moreover, if the errors decrease
exponentially, in the sense  that
 $\Delta(T^{\otimes n_\alpha},M_\alpha)\leq \mu e^{-\lambda n_\alpha}$
 ($\mu,\lambda\geq0$),
then they decrease exponentially for all $n$ with rate
\begin{equation}\label{gaprate}
 \liminf_{n\to\infty}\frac{-1}n
     \log\Delta(T^{\otimes n},\lfloor 2^{ cn }\rfloor)
    \geq\lambda.
\end{equation}
\end{lem}

\begin{proof}
Let us introduce the notation $c_+=\liminf_\alpha
(\log_2M_\alpha)/n_\alpha$, so $c<c_+$. We pick $\eta>0$ such that
$(1+\eta)c<c_+$. Then for sufficiently large $\alpha\geq\alpha_0$
we have $(n_{\alpha+1}/n_\alpha)\leq(1+\eta)$, and
$(\log_2M_\alpha/n_\alpha)\geq(1+\eta)c$. Now let $n\geq
n_{\alpha_0}$, and consider the unique index $\alpha$ such that
$n_\alpha\leq n\leq n_{\alpha+1}$. Then $n\leq (1+\eta)n_\alpha$
and
\begin{equation}
  \lfloor 2^{ cn }\rfloor
   \leq 2^{ cn }
   \leq 2^{ c(1+\eta)n_\alpha }
   \leq M_\alpha.
\end{equation}
Clearly, $\Delta(T^{\otimes n},M)$ decreases as $n$ increases,
because good coding becomes easier if we have more parallel
channels and increases with $M$, because if a coding scheme works
for an input Hilbert space $\HH_0$, it also works at least as well
for states supported on a lower dimensional subspace. Hence
$\Delta(T^{\otimes n},\lfloor 2^{ cn }\rfloor)\leq
\Delta(T^{\otimes n_\alpha},M_\alpha)\to0$. It follows that $c$ is
an admissible rate.

With the exponential bound on $\Delta$ we find similarly that
\begin{equation}
  \Delta(T^{\otimes n},\lfloor 2^{ cn }\rfloor)
   \leq \mu\;  e^{-\lambda n_\alpha}
   \leq \mu\; e^{-\lambda/(1+\eta) n},
\end{equation}
so that the liminf in (\ref{gaprate}) is $\geq\lambda/(1+\eta)$.
Since $\eta$ was arbitrary, we get the desired result.
\end{proof}

%%%%%%%%%%%%%%%%%%%%%%%%%%%%%%%%%%%%%%%%%%%%%%%%%%
\subsection{Elementary properties}
%%%%%%%%%%%%%%%%%%%%%%%%%%%%%%%%%%%%%%%%%%%%%%%%%%

To determine $Q(T)$ in terms of Definition \ref{def:1} is fairly difficult, because optimization problems
in spaces of exponentially fast growing dimensions are involved. This renders in particular each direct
numerical approach practically impossible. In the classical situation, i.e. if we transfer classical
information through a classical channel $\Phi$, we can define a capacity quantity $C(\Phi)$ in the same way as
above. An explicit calculation of $C(\Phi)$, however, can be reduced, according to Shannons ``noisy channel
coding theorem'' \cite{Shannon48}, to an optimization problem over a low dimensional space, which does not
involve the limit of inifinitely many parallel channels. A similar coding theorem for the quantum case
is not yet known -- this is the biggest open problem concerning channel capacities. 

Nevertheless, there are some special cases in which the capacity can be computed explicity. The most
relevant example is the ideal channel $\Id=\Id_{\scr{B}(\Bbb{C}^d)}$. If $d^n \geq M$ we can embed
$\Bbb{C}^M$ into $(\Bbb{C}^d)^{\otimes n}$, hence $\Delta(\Id^{\otimes n},M) = 0$ and we see that the rate $\log_2(d)$
can be achieved. Intuitively we expect that this is the best what can be done, because it is
impossible to embed a high- into a low-dimensional space. This intuition is in fact correct, i.e. we have
$Q(\Id) = \log_2(d)$ for the ideal channel. A precise proof of this statement is, however, not so easy as
it looks like and we skip the details here. Maybe the most easy approach is to use the quantity
$\log_2(\|\Theta T\|_\cb)$ (where $\Theta$ denotes the transposition), which is an upper bound on $Q(T)$
(cf. \cite{PRepQI} or \cite{Werner01}). The same idea can be used to show that the quantum
capacity of a classical channel, or more generally a channel $T$ which uses classical information at an
intermediate step,  is zero. This is a reformulation of the ``no classical teleportation theorem''
(cf. again  \cite{Werner01}). 

Another useful relation concerns the concatenation of two general channels $T_1$ and $T_2$: We transmit
quantum information first through  $T_1$ and then through $T_2$. It is reasonable to assume that the
capacity of the composition $T_2T_1$ can not be bigger than the capacity of the channel with the smallest
bandwidth. This conjecture is indeed true and known as the ``\emph{Bottleneck inequality}'':
  \begin{equation}  \label{eq:5}
    Q(T_2T_1) \leq \min \{ Q(T_1), Q(T_2)\}.
  \end{equation}
Alternatively we can use the two channels in parallel, i.e. we consider the tensor product $T_1 \otimes
T_2$. In this case the capacity of the resulting channel is at least as big as the sum of $Q(T_1)$ and
$Q(T_2)$, i.e. $Q$ is \emph{superadditive}:
\begin{equation} \label{eq:10}
  Q(T_1 \otimes T_2) \geq Q(T_1) + Q(T_2)
\end{equation}
(cf. \cite{PRepQI} for a proof of both statements). To decide whether $Q$ is even additive, i.e. whether
equality holds in (\ref{eq:10}), is another big open question about channel capacities.

% $Log: DDQCapQECC.tex,v $
% Revision 1.6  2002/05/03 12:04:41  michael
% Final version.
%
% Revision 1.5  2002/04/11 10:15:04  michael
% First complete draft. Several improvements are still necessary.
%
% Revision 1.4  2002/03/26 17:56:40  michael
% Many, many, many changes and additons by Reinhard.
%
% Revision 1.3  2002/03/06 15:30:12  michael
% Some references added. Knill and Laflamme stuff now complete.
%
% Revision 1.2  2002/03/05 19:00:12  michael
% Shortened the "channel" part and started with new Knill and Laflamme stuff.
%
% Revision 1.1  2002/03/05 11:03:18  michael
% Divided the project into several files. Intoduction added.
%
%
% $Id: DDQCapQECC.tex,v 1.6 2002/05/03 12:04:41 michael Exp $

%%%%%%%%%%%%%%%%%%%%%%%%%%%%%%%%%%%%%%%%%%%%%%%%%%%%%%%%%%%%%%%%%%%%%%%%%%%%%%%%%%%%%%%%%%%%%%%%%%%%
\section{Quantum error correction}
\label{sec:quant-error-corr}
%%%%%%%%%%%%%%%%%%%%%%%%%%%%%%%%%%%%%%%%%%%%%%%%%%%%%%%%%%%%%%%%%%%%%%%%%%%%%%%%%%%%%%%%%%%%%%%%%%%%

The definition of capacity requires that we correct errors in a
collection of $n$ parallel channels $T^{\otimes n}$. Here the
tensor product means that successive uses of the channel are
independent. For example, the physical system used as a carrier is
freshly prepared every time we use the channel. This independence
is important for error correcting schemes, because it prevents
errors happening on different channels to ``conspire''.

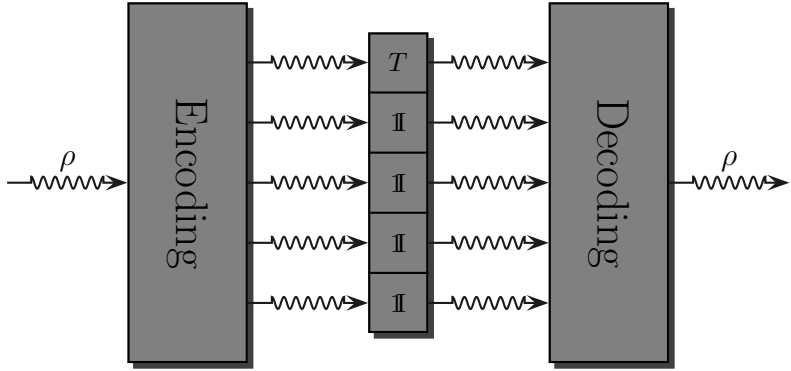
\begin{figure}[b]
  \begin{center}
    \begin{picture}(15,6)
      \psframe[fillcolor=meascolor,fillstyle=solid,shadow=true](3,0)(5,6)
      \psframe[fillcolor=meascolor,fillstyle=solid,shadow=true](7,0.5)(8,5.5)
      \psline{-}(7,1.5)(8,1.5)
      \psline{-}(7,2.5)(8,2.5)
      \psline{-}(7,3.5)(8,3.5)
      \psline{-}(7,4.5)(8,4.5)
      \psframe[fillcolor=meascolor,fillstyle=solid,shadow=true](10,0)(12,6)
      \pscoil[linecolor=qbitcolor,coilaspect=0,coilheight=1,coilwidth=.2]{->}(1,3)(3,3)
      \pscoil[linecolor=qbitcolor,coilaspect=0,coilheight=1,coilwidth=.2]{->}(12,3)(14,3)
      \pscoil[linecolor=qbitcolor,coilaspect=0,coilheight=1,coilwidth=.2]{->}(5,1)(7,1)
      \pscoil[linecolor=qbitcolor,coilaspect=0,coilheight=1,coilwidth=.2]{->}(5,2)(7,2)
      \pscoil[linecolor=qbitcolor,coilaspect=0,coilheight=1,coilwidth=.2]{->}(5,3)(7,3)
      \pscoil[linecolor=qbitcolor,coilaspect=0,coilheight=1,coilwidth=.2]{->}(5,4)(7,4)
      \pscoil[linecolor=qbitcolor,coilaspect=0,coilheight=1,coilwidth=.2]{->}(5,5)(7,5)
      \pscoil[linecolor=qbitcolor,coilaspect=0,coilheight=1,coilwidth=.2]{->}(8,1)(10,1)
      \pscoil[linecolor=qbitcolor,coilaspect=0,coilheight=1,coilwidth=.2]{->}(8,2)(10,2)
      \pscoil[linecolor=qbitcolor,coilaspect=0,coilheight=1,coilwidth=.2]{->}(8,3)(10,3)
      \pscoil[linecolor=qbitcolor,coilaspect=0,coilheight=1,coilwidth=.2]{->}(8,4)(10,4)
      \pscoil[linecolor=qbitcolor,coilaspect=0,coilheight=1,coilwidth=.2]{->}(8,5)(10,5)
      \rput(4,3){\LARGE \rotateright{Encoding}}
      \rput(11,3){\LARGE \rotateright{Decoding}}
      \rput(7.5,1){$\Bbb{1}$}
      \rput(7.5,2){$\Bbb{1}$}
      \rput(7.5,3){$\Bbb{1}$}
      \rput(7.5,4){$\Bbb{1}$}
      \rput(7.5,5){$T$}
      \rput[b](2,3.2){\large $\rho$}
      \rput[b](13,3.2){\large $\rho$}
    \end{picture}
    \caption{Five bit quantum code: Encoding one qubit into five and correcting one error.}
    \label{fig:five-bit-code}
  \end{center}
\end{figure}

Suggestive as it may be, quantum mechanics cautions us to be very
careful with this sort of language: just as we cannot assign
trajectories to quantum systems, it is problematic to speak about
errors `happening' in one channel, in a situation where we must
expect different classical pictures to `occur' in quantum
mechanical superposition. This is to be kept in mind, when we now
describe the theory of quantum error correcting codes in the sense
of Knill and Laflamme \cite{KnLafQECC}, which is very much based
on a classification of errors according the place where they
occur. For example, the coding/decoding pair $E,D$ will typically
have the property that $E(T_1\otimes T_2\otimes\cdots\otimes
T_n)D=\id$, whenever the number of positions at which
$T_i\neq\id$, i.e., the number of errors, is small (cf. Figure
\ref{fig:five-bit-code}).

\if1\withpics
\fi

In our presentation of the Knill-Laflamme Theory, we start from
the error corrector's dream, namely the situation in which \emph{all the
errors happen in another part of the system}, where we do not keep
any of the precious quantum information. This will help us to
characterize the structure of the kind of errors which such a
scheme may tolerate, or `correct'. Of course, the dream is just a
dream for the situation we are interested in: several parallel
channels, each of which may be affected by errors. But the
splitting of the system into subsystems, mathematically the
decomposition of the Hilbert space of the total system into a
tensor product is something we may change by a suitable unitary
transformation. This is then precisely the role of the encoding
and decoding operations. The Knill-Laflamme theory is precisely the
description of the situation where such a unitary, and hence a
coding/decoding scheme exists. Constructing such schemes, however,
is another matter, to which we will turn in the next section.

\subsection{An error corrector's dream}
So consider a system split into
$\scr{H}=\scr{H}_g\otimes\scr{H}_b$, where the indices $g$ and $b$
stand for `good' and `bad'. We prepare the system in a state
$\rho\otimes\KB\Omega\Omega$, where $\rho$ is the quantum state we
want to protect. Now come the errors in the form of a completely
positive map $T(A) = \sum_{i} F_i^* A F_i$. Then according to the
error corrector's dream, we would just have to discard the bad
system, and get the same state $\rho$ as before.

The hardest demands for realizing this come from pure states
$\rho=\KB\phi\phi$, because the only way that the restriction to
the good system can again be $\KB\phi\phi$ is that the state after
errors factorizes, i.e.
\begin{equation}\label{facterror}
  T_*(\KB{\phi\otimes\Omega}{\phi\otimes\Omega})
     =\sum_i\KB{F_i(\phi\otimes\Omega)}{F_i(\phi\otimes\Omega)}
     =\KB\phi\phi\otimes\sigma\;.
\end{equation}
This requires that
\begin{equation}\label{Fidream}
  F_i(\phi\otimes\Omega)=\phi\otimes \Phi_i\;,
\end{equation}
where $\Phi_i\in\scr{H}_b$ is some vector, which must be
independent of $\phi$ if such an equation is to hold for all
$\phi\in\scr{H}_g$. Conversely, condition (\ref{Fidream}) implies
(\ref{facterror}) for every pure state $\KB\phi\phi$ and, by
convex combination, for every state $\rho$.

Two remarks are in order. Firstly, we have \emph{not} required
that $F_i=\idty\otimes F_i'$. This would be equivalent to
demanding that this scheme works with every $\Omega$, or indeed
with every (possibly mixed) initial state of the bad system. This
would be much too strong for a useful theory of codes. So later on
we must insist on a proper initialization of the bad subsystem by
a suitable encoding. Secondly, if we have the condition
(\ref{Fidream}) for the Kraus operators of some channel $T$, then
it also holds for all channels whose Kraus operators can be
written as linear combinations of the $F_i$. In other words, the
``set of correctible errors'' is naturally identified with the
vector space of operators $F$ such that there is a vector
$\Phi\in\scr{H}_b$ with  $F(\phi\otimes\Omega)=\phi\otimes\Phi$
for all $\phi\in\scr{H}_g$. This space will be called the
\emph{maximal error space} of the coding scheme, and will be
denoted by $\scr E_{\rm max}$. Usually, a code is designed for a
given error space $\scr E$. Then the statement that these given
errors are corrected simply becomes $\scr E\subset\scr E_{\rm
max}$. The key observation, however, is that the space of errors
is a vector space in a natural way, i.e., if we can correct two
types of errors, then we can also correct their
\emph{superposition}.

\subsection{Realizing the dream by unitary transformation}
Let us now consider the situation in which we want to send states
of a small system with Hilbert space ${\scr H}_1$ through a
channel $T:{\scr B}({\scr H}_2)\to{\scr B}({\scr H}_2)$. The Kraus
operators of $T$ lie in an error space $\scr{E}\subset{\scr
B}({\scr H}_2)$, which we assume to be given. No more assumptions
will be made about $T$. Our task is now to devise coding $E$ and
decoding $D$ so that $ETD$ is the identity on ${\scr B}({\scr
H}_1)$.

The idea is to realize the error corrector's dream by suitable
encoding. The `good' space in that scenario is, of course, the
space ${\scr H}_1$. We are looking for a way to write ${\scr
H}_2\cong {\scr H}_1\otimes{\scr H}_b$. Actually, an isomorphism
may be asking too much, and we look for an isometry $U:{\scr
H}_1\otimes{\scr H}_b\to{\scr H}_2$. The encoding, written best in
the Schr{\"o}dinger picture, is tensoring with an initial state
$\Omega$ as before, but now with an additional twist by $U$:
\begin{equation}\label{encodeU}
  E_*(\rho) = U(\rho \otimes \kb{\Omega})U^*\;.
\end{equation}
The decoding operation $D$ is again taking the partial trace over
the bad space ${\scr H}_b$, after reversing of $U$. Since $U$ is
only an isometry and not necessarily unitary we need an additional
term to make $D$ unit preserving. The whole operation is is best
written in the Heisenberg picture:
\begin{equation}
  D(X) = U (X \otimes \idty) U^*
         +  \tr(\rho_0 X) (\Bbb{1} - UU^*)\;,
\end{equation}
 where $\rho_0$ is an arbitrary density operator.
These transformations are successful, if the error space
(transformed by $U$) behaves as before, i.e., if for all
$F\in{\scr E}$ there are vectors $\Phi(F)\in{\scr H}_b$ such that,
for all $\phi\in{\scr H}_1$
\begin{equation} \label{eq:4}
  F U(\phi \otimes \Omega) = U(\phi \otimes \Phi(F))
\end{equation}
holds. This equation describes precisely the elements $F\in\scr
E_{\rm max}$ of the maximal error space.

To check that we really have $ETD = \Id$ for any channel
$T(A)=\sum_iF_i^*AF_i$  with $F_i\in{\scr E_{\rm max}}$, it
suffices to consider pure input states $\kb\phi$, and the measurement of an
arbitrary observable $X$ at the output:
\begin{align}
  \tr \bigl[ \kb{\phi} ETD(X)\bigr]
    &= \sum_{i} \tr\bigl[ U \kb{\phi\otimes \Omega}U^* F_i
                          U(X \otimes\Bbb{1}) U^* F_i\bigr]  \nonumber\\
  &= \sum_{i} \tr \bigl[ \kb{\phi\otimes \Phi(F_i)} X \otimes \Bbb{1}\bigr]
      \nonumber\\
  &= \langle\phi, X \phi\rangle \sum_{i}\|\Phi(F_i)\|^2
  = \langle\phi,X\phi\rangle.
\end{align}
In the last equation we have used that
$\sum_{i}\|\Phi(F_i)\|^2=1$, since $E,T$, and $D$ each map
$\idty$ to $\idty$.

\subsection{The Knill-Laflamme condition}\label{sec:KniLa}
The encoding $E$ defined in Equation (\ref{encodeU}) is of the
form $E_*(\rho)=V\rho V^*$ with the \emph{encoding isometry}
$V:\scr{H}_1 \to \scr{H}_2$ given by
\begin{equation}\label{encIso}
  V\phi = U(\phi \otimes \Omega)\;.
\end{equation}
 If we just know this isometry and the error space we can
reconstruct the whole structure, including the decomposition
${\scr H}_2={\scr H}_1\otimes{\scr H}_b\oplus(\idty-UU^*){\scr
H}_2$, and hence the decoding operation $D$. A necessary condition
for this, first established by Knill and Laflamme \cite{KnLafQECC}, is
that, for arbitrary $\phi_1,\phi_2\in{\scr H}_1$ and error
operators $F_1,F_2\in{\scr E}$:
\begin{equation} \label{KLaF}
  \langle V\phi_1, F_1^*F_2 V\phi_2\rangle
   = \langle\phi_1, \phi_2\rangle \omega(F_1^*F_2)
\end{equation}
holds with some numbers $\omega(F_1^*F_2)$ independent of
$\phi_1,\phi_2$.  Indeed, from (\ref{eq:4}) we immediately get
this equation with
$\omega(F_1^*F_2)=\langle\Phi(F_1),\Phi(F_2)\rangle$. Conversely,
if the Knill-Laflamme condition (\ref{KLaF}) holds, the numbers
$\omega(F_1^*F_2)$ serve as a (possibly degenerate) scalar product
on ${\scr E}$, which upon completion becomes the `bad space'
${\scr H}_b$, such that $F\in\scr E$ is identified with a Hilbert
space vector $\Phi(F)$. The operator $U:\phi\otimes\Phi(F)=FV\phi$
is then an isometry, as used at the beginning of this section. To
conclude, the Knill-Laflamme condition is necessary and sufficient
for the existence of a decoding operation. Its main virtue is that
we can use it without having to construct the decoding explicitly.

\subsection{Example: Localized errors}
Let us come back to the problem we are addressing in this paper.
In that case the space $\HH_2$ is the $n$-fold tensor product of
the system $\HH$ on which the noisy channels under consideration
act. We say that a coding isometry $V:\HH_1\to\HH^{\otimes n}$
\emph{corrects $f$ errors}, if it satisfies the Knill-Laflamme
condition (\ref{KLaF}) for the error space $\scr E_f$ spanned
linearly by  all operators of the kind $X_1\otimes
X_2\otimes\cdots\otimes X_n$, where at most $f$ places we have a
tensor factor $X_i\neq\idty$.

When $F_1$ and $F_2$ are both supported on at most $f$ sites, the
product $F_1^*F_2$, which appears in the Knill-Laflamme condition
involves $2f$ sites. Therefore we can paraphrase the condition by
saying that
\begin{equation} \label{KLaF2}
  \langle V\phi_1, X V\phi_2\rangle
   = \langle\phi_1, \phi_2\rangle \omega(X)
\end{equation}
for $X\in{\scr E}_{2f}$. From Kraus operators in $\scr E_f$ we can
build arbitrary channels of the kind $T=T_1\otimes
T_2\otimes\cdots\otimes T_n$, where at most $f$ of the tensor
factors $T_i$ are channels different from $\id$. We will use this
in the form that $E(R_1\otimes R_2\otimes\cdots\otimes R_n)D=0$,
whenever at most $f$ tensor factors are $R_i\neq\id$, and at least
one of them is a difference of two channels.

There are several ways to construct error correcting codes of this
type (see e.g. \cite{GotQECC,CRSSQECC,AsKnQECC}). Most appropriate
for our purposes is the scheme proposed in \cite{Schlingel}, which
is quite easy to describe and admits a simple way to check the
error correction condition. This will be the subject of the next
section.

%%%%%%%%%%%%%%%%%%%%%%%%%%%%%%%%%%%%%%%%%%%%%%%%%%%%%%%%%%%%%%%%%%%%%%%%%%%%%%%%%%%%%%%%%%%%%%%%%%%%
\section{Graph Codes}
\label{sec:graph-codes}
%%%%%%%%%%%%%%%%%%%%%%%%%%%%%%%%%%%%%%%%%%%%%%%%%%%%%%%%%%%%%%%%%%%%%%%%%%%%%%%%%%%%%%%%%%%%%%%%%%%%

The general scheme of graph codes works not just for qubits, but
for any dimension $d$ of one site spaces.  The code will have some
number $m$ of input systems, which we label by a set $X$, and,
similarly $n$ output systems, labeled by a set $Y$. The Hilbert
space  of the system with label $x\in X\cup Y$ will be denoted by
$\HH_x$ although all these are isomorphic to $\Cx^d$, and are
equipped with a special basis $\ket{j_x}$, where $j_x\in\Zd$ is an
integer taken modulo $d$. As a convenient shorthand, we write
$j_X$ for a tuple of $j_x\in\Zd$, specified for every $x\in X$.
Thus the $\ket{j_X}$ form a basis of the input space
$\HH_X=\bigotimes_{x\in X}\HH_x$ of the code. An operator $F$,
say, on the output space will be called \emph{localized} on a
subset $Z\subset Y$ of systems, if it is some operator on
$\bigotimes_{y\in Z}\HH_y$, tensored with the identity operators
of the remaining sites.

\if1\withpics
\begin{wrapfigure}{l}{5.5cm}
  \psset{unit=3mm}
    \begin{pspicture}(16,8)
      \rput(4,4){
        \SpecialCoor
        \cnode[fillstyle=solid,fillcolor=black](0,0){0.2}{A}
        \cnode[fillstyle=solid,fillcolor=black](3.5;90){0.2}{B}
        \cnode[fillstyle=solid,fillcolor=black](3.5;18){0.2}{C}
        \cnode[fillstyle=solid,fillcolor=black](3.5;-54){0.2}{D}
        \cnode[fillstyle=solid,fillcolor=black](3.5;162){0.2}{E}
        \cnode[fillstyle=solid,fillcolor=black](3.5;232){0.2}{F}
        \psset{linewidth=1pt}
        \ncline{-}{A}{B}
        \ncline{-}{A}{C}
        \ncline{-}{A}{D}
        \ncline{-}{A}{E}
        \ncline{-}{A}{F}
        \ncline{-}{B}{C}
        \ncline{-}{C}{D}
        \ncline{-}{B}{E}
        \ncline{-}{E}{F}
        \ncline{-}{F}{D}
        \NormalCoor
        }
      \rput(12,4){
        \cnode[fillstyle=solid,fillcolor=black](-3.5,-2){0.2}{A}
        \cnode[fillstyle=solid,fillcolor=black](-3.5,2){0.2}{B}
        \cnode[fillstyle=solid,fillcolor=black](3.5,-2){0.2}{C}
        \cnode[fillstyle=solid,fillcolor=black](3.5,2){0.2}{D}
        \cnode[fillstyle=solid,fillcolor=black](-2.5,0){0.2}{E}
        \cnode[fillstyle=solid,fillcolor=black](2.5,0){0.2}{F}
        \psset{linewidth=1pt}
        \ncline{-}{A}{B}
        \ncline{-}{B}{D}
        \ncline{-}{D}{C}
        \ncline{-}{C}{A}
        \ncline{-}{A}{E}
        \ncline{-}{B}{E}
        \ncline{-}{C}{F}
        \ncline{-}{D}{F}
        \ncline{-}{E}{F}
     }
    \end{pspicture}
  \caption{Two graph codes.}
  \label{fig:graphs}
\end{wrapfigure}
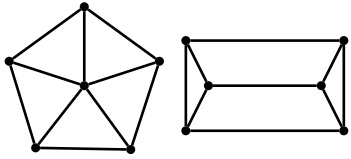
\fi
The main ingredient of the code construction is now an undirected
graph with vertices $X\cup Y$. The links of the graph are given by
the adjacency matrix, which we will denote by $\Gamma$. When we
have $|X|=m$ input vertices and $|Y|=n$ output vertices, this is
an $(n+m)\times(n+m)$ matrix with $\Gamma_{xy} = 1$ if node $x$
and $y$ are linked and $\Gamma_{xy} = 0$ otherwise. We do allow
multiple edges, so the entries of $\Gamma$ will in general be
integers, which can also be taken modulo $d$. It is convenient to
exclude self-linked vertices, so we always take $\Gamma_{xx}=0$.

The graph determines an operator $V=V_\Gamma: \scr{H}_X \to
\scr{H}_Y$ by the formula
\begin{equation} \label{eq:27}
  \langle j_Y | V_\Gamma | j_X\rangle
    = d^{-n/2}\exp\left(\frac{i\pi}{d}\ j_{X\cup Y}
             \cdot \Gamma\cdot j_{X\cup Y}\right),
\end{equation}
where the exponent contains the matrix element of $\Gamma$
\begin{equation}\label{graphexponent}
  j_{X\cup Y}\cdot \Gamma\cdot j_{X\cup Y}
     =\sum_{x,y\in X\cup Y}j_x\Gamma_{xy}j_y\;.
\end{equation}
Because $\Gamma$ is symmetric, every term in this sum appears
twice, hence adding a multiple of $d$ to any $j_x$ or
$\Gamma_{xy}$ will change the exponent in (\ref{eq:27}) by a
multiple of $2\pi$, and thus will not change $V_\Gamma$.

The error correcting properties of $V_\Gamma$ are summarized in
the following result \cite{Schlingel}. It is just the
Knill-Laflamme condition with a special expression for the form
$\omega$,  for error operators such that $F_1^*F_2$ is localized
on a set $Z$.

\begin{prop}
Let $\Gamma$ be a graph, i.e., a symmetric matrix with entries
$\Gamma_{xy}\in\Zd$, for $x,y\in(X\cup Y)$. Consider a subset
$Z\subset Y$, and suppose that the\break
 $(Y\setminus Z)\times(X\cup Z)$-submatrix of $\Gamma$ is
non-singular, i.e.,
\begin{equation}\label{Zcorrects}
  \forall_{y\in Y\setminus Z}\ \sum_{x\in X\cup Z}\Gamma_{yx}h_x\equiv0
       \quad\text{implies\ }
  \forall_{x\in X\cup Z}\ h_x\equiv0
\end{equation}
where congruences are $\mod d$. Then, for every operator
$F\in\B(\HH_Y)$ localized on $Z$, we have
\begin{equation}\label{KL4Z}
  V_\Gamma^*FV_\Gamma
  =d^{-n}\tr(F)\idty_X
\end{equation}
\end{prop}

\begin{proof}
It will be helpful to use the notation for collections of
variables, already present in (\ref{graphexponent}) more
systematically: for any subset $W\subset X\cup Y$ we write $j_W$
for the collection of variables $j_y$ with $y\in W$. The
Kronecker-Delta $\delta(j_W)$ is defined to be zero if for any
$y\in W$ $j_y\neq0$, and one otherwise. By
$j_W\cdot\Gamma_{WW'}\cdot k_{W'}$ we mean the suitably restricted
sum, i.e., $\sum_{x\in W,y\in W'}j_x\Gamma_{xy}k_y$. The important
sets to which we apply this notation are $X'=(X\cup Z)$ and
$Y'=Y\setminus Z$. In particular, the condition on $\Gamma$ can be
written as $\Gamma_{Y'X'}j_{X'}=0\implies j_{X'}=0$.

Consider now the matrix element
\begin{eqnarray}\label{bigsum4graph}
 \langle j_X\vert V_\Gamma^*FV_\Gamma\vert k_X\rangle
 &=&
 \sum_{j_Y,k_Y} \langle j_X\vert V_\Gamma^*\vert j_Y\rangle
                \langle j_Y\vert F\vert k_Y\rangle
                \langle k_Y\vert V_\Gamma\vert k_X\rangle
 \\&=&d^{-n}
 \sum_{j_Y,k_Y} e^{\frac{i\pi}d\Bigl(
      k_{X\cup Y}\cdot \Gamma\cdot k_{X\cup Y}-j_{X\cup Y}\cdot \Gamma\cdot j_{X\cup Y}
      \Bigr)}\
                \langle j_Y\vert F\vert k_Y\rangle
\nonumber
\end{eqnarray}
Since $F$ is localized on $Z$, the matrix element contains a
factor $\delta_{j_y,k_y}$ for every $y\in Y\setminus Z=Y'$, so we
can write $\langle j_Y\vert F\vert k_Y\rangle=\langle j_Z\vert
F\vert k_Z\rangle\delta(j_{Y'}-k_{Y'})$. Therefore we can compute
the sum (\ref{bigsum4graph}) in stages:
\begin{equation}\label{oeae}
   \langle j_X\vert V_\Gamma^*FV_\Gamma\vert k_X\rangle
      =\sum_{j_Z,k_Z}\langle j_Z\vert F\vert k_Z\rangle
      S(j_{X'},k_{X'})\;,
\end{equation}
where $S(j_{X'},k_{X'})$ is the sum over the $Y'$-variables,
which, of course, still depends on the input variables $j_X,k_X$
and the variables $j_Z,k_Z$ at the error positions:
\begin{equation}
  S(j_{X'},k_{X'})
     =d^{-n}\sum_{j_{Y'},k_{Y'}}\delta({j_{Y'}-k_{Y'}})
      e^{\frac{i\pi}d\Bigl(k_{X\cup Y}\cdot \Gamma\cdot k_{X\cup Y}
                          -j_{X\cup Y}\cdot \Gamma\cdot j_{X\cup Y}
      \Bigr)}
\end{equation}
The sums in the exponent can each be split into four parts
according to the decomposition $X'$ vs.\ $Y'$. The terms involving
$\Gamma_{Y'Y'}$ cancel because $k_{Y'}=j_{Y'}$. The terms
involving $\Gamma_{X'Y'}$ and $\Gamma_{Y'X'}$ are equal because
$\Gamma$ is symmetric, and together give
$2j_{Y'}\cdot\Gamma_{Y'X'}\cdot(k_{X'}-j_{X'})$. The
$\Gamma_{X'X'}$ remain unchanged, but only give a phase factor
independent of the summation variables. Hence
\begin{eqnarray}
   S(j_{X'},k_{X'})
   &=&d^{-n}e^{\frac{i\pi}d\bigl(k_{X'}\cdot \Gamma\cdot k_{X'}
                          -j_{X'}\cdot \Gamma\cdot j_{X'}\bigr)}
      \sum_{j_{Y'}}e^{\frac{2\pi i}dj_{Y'}\cdot\Gamma_{Y'X'}\cdot(k_{X'}-j_{X'})}
   \nonumber\\
   &=&d^{-n}e^{\frac{i\pi}d\bigl(k_{X'}\cdot \Gamma\cdot k_{X'}
                          -j_{X'}\cdot \Gamma\cdot j_{X'}\bigr)}
      d^{|Y'|} \; \delta(\Gamma_{Y'X'}\cdot(k_{X'}-j_{X'}))
    \nonumber\\
   &=&d^{-n+|Y'|}e^{\frac{i\pi}d\bigl(k_{X'}\cdot \Gamma\cdot k_{X'}
                          -j_{X'}\cdot \Gamma\cdot j_{X'}\bigr)}
                          \delta(k_{X'}-j_{X'})
     \nonumber\\
   &=&d^{-n+|Y'|} \delta(k_{X'}-j_{X'})\;.
\end{eqnarray}
Here we used at the first equation that the sum is a product of
geometric series as they appear in discrete Fourier transforms. At
the second equality the main condition of the Proposition enters:
if $\sum_{x\in X'}\Gamma_{yx}\cdot(k_{x}-j_{x})$ vanishes for all
$y\in Y'$ as required by the delta-function then (and only then)
the vector $k_{X'}-j_{X'}$ must vanish. But then the two terms in
the exponent of the phase factor also cancel.

Inserting this result into (\ref{oeae}), and using that
$\delta(h_{X'})=\delta(h_X)\delta(h_Z)$, we find
\begin{eqnarray}
  \langle j_X\vert V_\Gamma^*FV_\Gamma\vert k_X\rangle
      &=& \delta(j_X-k_X)\  d^{-n+|Y'|}\sum_{j_Z}\langle j_Z\vert F\vert j_Z\rangle
      \nonumber\\
      &=& \delta(j_X-k_X)\ d^{-n} \sum_{j_Y}\langle j_Y\vert F\vert j_Y\rangle
      \nonumber
\end{eqnarray}
Here  the error operator is considered in the first line as an
operator on $\HH_Z$, and as an operator on $\HH_Y$ in the second
line, by tensoring it with $\idty_{Y'}$. This cancels the
dimension factor $d^{|Y'|}$
\end{proof}

All that is left to get an error correcting code is to ensure that
the conditions of this Proposition are satisfied sufficiently
often. This is evident from combining the above Proposition with
the example at the end of Section~\ref{sec:KniLa}.

\begin{kor}
Let $\Gamma$ be a graph as in the previous Proposition, and
suppose that the $(Y\setminus Z)\times(X\cup Z)$-submatrix of
$\Gamma$ is non-singular for \emph{all} $Z\subset Y$ with up to
$2f$ elements. Then the code associated to $\Gamma$ corrects $f$
errors.
\end{kor}

Two particular examples (which are equivalent!) are given in
Figure \ref{fig:graphs}. In both cases we have $N=1$, $M=5$ and
$K=1$ i.e. one input node, which can be chosen arbitrarily, five
output nodes and the corresponding codes correct one error.

%%% Local Variables: 
%%% mode: latex
%%% TeX-master: "DDQCap"
%%% End: 
%%%%%%%%%%%%%%%%%%%%%%%%%%%%%%%%%%%%%%%%%%%%%%%%%%%%%%%%%%%%%%%%%%%%%%%%%%%%%%%%%%%%%%%%%%%%%%%%%%%%
\section{Discrete to continuous error model}
\label{sec:discr-cont-error}
%%%%%%%%%%%%%%%%%%%%%%%%%%%%%%%%%%%%%%%%%%%%%%%%%%%%%%%%%%%%%%%%%%%%%%%%%%%%%%%%%%%%%%%%%%%%%%%%%%%%

The discrete error correction scheme described in the last section
is not really designed to correct {\em small} errors: it corrects
{\em rare} errors in multiple applications of the channel. A
typical example of a small (but not rare) error is a small unitary
rotation, $T(X)=U^*XU$. Then $\cbnorm{T-\id}$ can be small, but
since the same small error happens to each of the parallel
channels in $T^{\otimes n}$, the error syndromes of discrete error
correction at first sight do not seem to be appropriate at all.
Nevertheless, the discrete theory can be applied, and this is the
content of the following Proposition. It is the appropriate
formulation of ``reducing the order of errors from $\varepsilon$
to $\varepsilon^{f+1}$''.

\begin{prop} \label{Pinomi}
Let $T:\B(\HH)\to\B(\HH)$ be a channel, and let $E,D$ be encoding
and decoding channels for coding $m$ systems into $n$ systems.
Suppose that this coding scheme corrects $f$ errors, and that
\begin{equation} \label{eq:11}
  \cbnorm{T-\id}\leq(f+1)/(n-f-1).  
\end{equation}
Then
\begin{equation}\label{binbound}
  \cbnorm{ET^{\otimes n}D-\id}
   \leq\cbnorm{T-\id}^{f+1}\;2^{nH_2((f+1)/n)}\;,
\end{equation}
where $H_2(r)=-r\log_2 r-(1-r)\log_2(1-r)$ denotes the Shannon
entropy of the probability distribution $(r,1-r)$.
\end{prop}

\begin{proof}
Into $ET^{\otimes n}D$, we insert the decomposition
$T=\id+(T-\id)$ and expand the product. This gives $2^n$ terms,
containing tensor products with some number, say $k$, of tensor
factors $(T-\id)$ and tensor factors $\id$ on the remaining
$(n-k)$ sites. Now when $k\leq f$, the error correction property
makes the term zero. Terms with $k>f$ we estimate by
$\cbnorm{T-\id}^k$. Collecting terms we get
\begin{equation}
  \cbnorm{ET^{\otimes n}D-\id}
   \leq\sum_{k=f+1}^n{n\choose k}\cbnorm{T-\id}^k\;.
\end{equation}
The rest then follows from the next Lemma (with $r=(f+1)/n$). It
treats the exponential growth in $n$ for truncated binomial sums.

\begin{lem} \label{lem:3} Let  $0\leq r\leq1$ and $a>0$ such that
$a\leq r/(1-r)$. Then, for all integers $n$:
\begin{equation}\label{bintrunc}
 \frac1n \log\left(\sum_{k=rn}^n{n\choose k}a^k\right)
   \leq \log\bigl(a^r)+H_2(r)\;.
\end{equation}
\end{lem}

\begin{proof} For $\lambda>0$ we can estimate the step function by
an exponential, and get
\begin{eqnarray}
  \sum_{k=rn}^n{n\choose k}a^k
   &\leq&\sum_{k=0}^n{n\choose k}a^k e^{\lambda(k-rn)}
           \nonumber\\
   &=&e^{-\lambda rn}\bigl(1+ae^{\lambda}\bigr)^n= M(\lambda)^n
\end{eqnarray}
with $M(\lambda)=e^{-\lambda r}\bigl(1+ae^{\lambda}\bigr)$. The
minimum over all real $\lambda$ is attained at $ae^{\lambda_{\rm
min}} =r/(1-r)$. We get $\lambda_{\rm min}\geq0$ precisely when
the conditions of the Lemma are satisfied, in which case the bound
is computed by evaluating $M(\lambda)$.
\end{proof}
\end{proof}

Suppose now that we find a family of coding schemes with
$n,m\to\infty$ with fixed rate $r\approx(m/n)$ of inputs per
output, and a certain fraction  $f/n\approx\varepsilon$ of errors
being corrected. Then we can apply the Proposition and find that
the errors can be estimated above by
\begin{equation}\label{errorexpt}
  \Delta\left(T^{\otimes n},d^{m}\right)
    \leq \left(2^{H_2(\varepsilon)}\;
         \cbnorm{T-\id}^\varepsilon \right)^n\;,
\end{equation}
where $d$ is the Hilbert space dimension of each input system.
This goes to zero, and even exponentially to zero, as soon as the
expression in parentheses is $<1$. This will be the case whenever
$\cbnorm{T-\id}$ is small enough, or, more precisely,
\begin{equation}\label{normsmall}
   \cbnorm{T-\id}\leq\; 2^{-H_2(\varepsilon)/\varepsilon}.
\end{equation}
Note in addition that we have for all $n \in \Bbb{N}$
\begin{equation}
  2^{H_2(\varepsilon)/\varepsilon} < \frac{\epsilon - \frac{1}{n}}{1 - \epsilon + \frac{1}{n}}.
\end{equation}
Hence the bound from Equation (\ref{eq:11}) is implied by
(\ref{normsmall}). 

\begin{figure}[h]
  \begin{center}
    \begin{pspicture}(15,10)
    \rput(7.5,5){\includegraphics[scale=0.8]{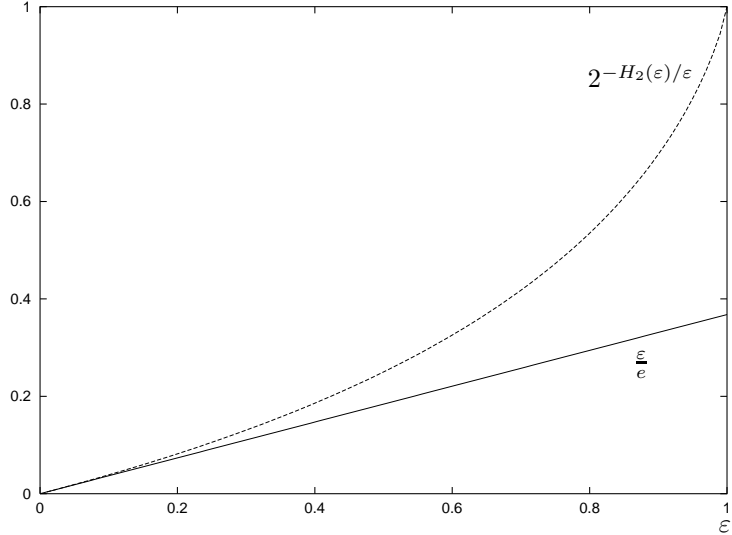}}
    \rput[r](13.5,0.5){$\epsilon$}
    \rput(12,8){$2^{-H_2(\epsilon)/\epsilon}$}
    \rput(12,3.2){$\frac{\epsilon}{e}$}
  \end{pspicture} 
    \caption{The two bounds from Equation (\ref{simplebound}) plotted
      as a function of $\epsilon$.} 
    \label{fig:bounds1}
  \end{center}
\end{figure}

The function appearing on the right hand side of (\ref{normsmall})
looks rather complicated, so we will often replace it by a simpler one, namely
\begin{equation}\label{simplebound}
  \frac\varepsilon e \leq 2^{-H_2(\varepsilon)/\varepsilon}\;,
\end{equation}
where $e$ is the base of natural logarithms; cf. Figure
\ref{fig:bounds1}. The proof of this inequality is left to the reader
as exercise in logarithms. The bound is very good (exact to first
order) in the range of small $\varepsilon$, in which we are most interested
anyhow. In any case, from $\cbnorm{T-\id}\leq\varepsilon/e$ we can draw the same
conclusion as from (\ref{normsmall}): exponentially decreasing
errors, provided we can actually find code families correcting a
fraction $\epsilon$ of errors. This will be the aim of the next
section.

%%%%%%%%%%%%%%%%%%%%%%%%%%%%%%%%%%%%%%%%%%%%%%%%%%%%%%%%%%%%%%%%%%%%%%%%%%%%%%%%%%%%%%%%%%%%%%%%%%%%
\section{Coding by random graphs}
\label{sec:coding-random-graphs}
%%%%%%%%%%%%%%%%%%%%%%%%%%%%%%%%%%%%%%%%%%%%%%%%%%%%%%%%%%%%%%%%%%%%%%%%%%%%%%%%%%%%%%%%%%%%%%%%%%%%

Our aim in this section is to apply the theory of graph codes to
construct a family of codes with positive rate. It is not so easy
to construct such families explicitly. However, if we are only
interested in existence, and do not attempt to get the best
possible rates, we can use a simple argument, which shows not only
the existence of codes  correcting  a certain fraction of errors,
but even that ``typical graph codes'' for sufficiently large
numbers of inputs and outputs have this property. Here ``typical''
is in the sense of the probability distribution, defined by simply
setting the edges of the graph independently, and each according
to the uniform distribution of the possible values of the
adjacency matrix. For the random method to work we need the
dimension of the underlying one site Hilbert space to be a prime
number. This curious condition is most likely an artefact of our
method, and will be removed later on.

We have seen that a graph code corrects many errors if certain
submatrices of the adjacency matrix have maximal rank. Therefore
we need the following Lemma.

\begin{lem} Let $d$ be a prime, $M<N$ integers and let $X$ be an
$N\times M$-matrix with independent and uniformly distributed
entries in $\Zd$. Then $X$ is singular over the field $\Zd$ with
probability at most $d^{-(N-M)}$.
\end{lem}

\begin{proof} The sum of independent uniformly distributed random
variables in $\Zd$ is again uniformly distributed. Moreover, since
$d$ is prime, this distribution is invariant under multiplication
by non-zero factors. Hence if $x_j\in\Zd$ ($j=1,\ldots,N$)are
independent and uniformly distributed, and $\phi_j\in\Zd$ are
non-random constants, not of all of which are zero,
$\sum_{j=1}^Nx_j\phi_j$ is uniformly distributed. Hence, for a
fixed vector $\phi\in\Zd^M$, the $N$ components
$(X\phi)_k=\sum_{j=1}^MX_{kj}\phi_j$ are independent uniformly
distributed random variables. Hence the probability for $X\phi=0$
for some fixed $\phi\neq0$ is $d^{-N}$. Since there are $d^M-1$
vectors $\phi$ to be tested, the probability for {\it some} $\phi$
to yield $X\phi=0$ is at most $d^{M-N}$. \end{proof}

\begin{prop} \label{prop:2} Let $d$ be a prime, and let $\Gamma$ be a symmetric 
$(n+m)\times(n+m)$-matrix with entries in $\Zd$, chosen at random
such that $\Gamma_{kk}=0$ and that the $\Gamma_{kj}$ with $k>j$
are independent and uniformly distributed. Let $P$ be the
probability for the corresponding graph code {\em not} to correct
$f$ errors (with $2f < n$). Then
\begin{equation}\label{Pnotcorrect}
  \frac1n\log P
    \leq \Bigr(\frac mn+\frac{4f}n-1\Bigl)\log d
        +H_2\Bigl(\frac{2f}n\Bigr)\;.
\end{equation}
\end{prop}

\begin{proof} Each error configuration is an $2f$-element subset of
the $n$ output nodes. According to Proposition ... we have to
decide, whether the corresponding $(n-2f)\times(m+2f)$-submatrix
of $\Gamma$, connecting input and error positions with the
remaining output positions, is singular or not. Since this
submatrix contains no pairs $\Gamma_{ij},\Gamma_{ji}$, its entries
are independent and satisfy the conditions of the previous Lemma.
Hence the probability that a particular configuration of $e$
errors goes uncorrected is at most $d^{(m+2f)-(n-2f)}$. Since
there are ${n\choose 2f}$ possible error configurations among the
outputs, we can estimate the probability of any $2f$ site error
configuration to be undetected as less than ${n\choose
2f}d^{m-n+4f}$. Using Lemma \ref{lem:3} we can estimate the
binomial as $\log{n\choose 2f}\leq n H_2(2f/n)$, which leads to
the bound stated.
\end{proof}

In particular, if the right hand side of the inequality in
(\ref{Pnotcorrect}) is negative, we get $P<1$, so that there must
be at least one matrix $\Gamma$ correcting $f$ errors. The crucial
point is that this observation does not depend on $n$, but only on
the rate-like parameters $m/n$ and $f/n$. Let us make this
behaviour a Definition:

\begin{defi} Let $d$ be an integer. Then we say a pair $(\mu,\epsilon)$
consisting of a \emph{coding rate} $\mu$ and an \emph{error rate}
$\epsilon$ is \emph{achievable}, if for every $n$ we can find an
encoding $E$ of $\lceil \mu n\rceil$ $d$-level systems into $n$
$d$-level systems correcting $\lfloor \epsilon n\rfloor$ errors.
\end{defi}

Then we can paraphrase the last proposition as saying that all
pairs $(\mu,\epsilon)$ with
\begin{equation}\label{eq:8}
  (1-\mu-4\epsilon)\log_2 d> H_2(2\epsilon)
\end{equation}
are achievable. This is all the input we need for the next
section, although a better coding scheme, giving larger $\mu$ or larger
$\epsilon$ would also improve the rate estimates proved there. Such improvements are indeed possible. E.g. for
the qubit case ($d=2)$ it is shown in \cite{CRSSQECC} that there is allways a code which saturates the
\emph{quantum Gilbert-Varshamov bound} $(1 - \mu - 2 \epsilon \log_2(3)) > H_2(2\epsilon)$ which is slightly better than
our result.

But there are also known limitations, particularly the so-called
\emph{Hamming bound}. This is a simple dimension counting
argument, based on the error correctors dream: Assuming that the
scalar product $(F,G)\mapsto\omega(F^*G)$ on the error space $\scr
E$ is non-degenerate, the dimension of the ``bad space'' is the
same as the dimension of the error space. Hence with the notations
of Section~\ref{sec:quant-error-corr} we expect
 $\dim\HH_0\cdot\dim{\scr E}\leq\dim\HH_2$. We now take $m$ input
systems and $n$ output systems of dimension $d$ each, so that
$\dim\HH_1=d^m$ and $\dim\HH_2=d^n$. For the space of errors
happening at at most $f$ places we introduce a basis s follows: at
each site we choose a basis of $\scr B(\HH)$ consisting of $d^2-1$
operators plus the identity. Then a basis of $\scr E$ is given by
all tensor products with basis elements $\neq\idty$ placed at
$j\leq f$ sites. Hence
 $\dim\scr E=\sum_{j\leq f}{n\choose j}(d^2-1)^j$.
For large $n$ we estimate this as in Lemma~\ref{Pinomi} as
$\log\dim\scr E\approx(f/n)\log_2(d^2-1)+H_2(f/n)$. Hence the
Hamming bound becomes
\begin{equation}\label{Hamming}
   \frac mn\log_2d +H_2(\epsilon)+\frac fn\log_2(d^2-1)\leq \log_2d
\end{equation}
which (with $d^2 \gg 1$) is just (\ref{eq:8}) with a factor $1/2$ on all
errors.
\begin{figure}[t]
  \begin{center}
    \begin{pspicture}(15,10)
    \rput(7.5,5){\includegraphics[scale=0.8]{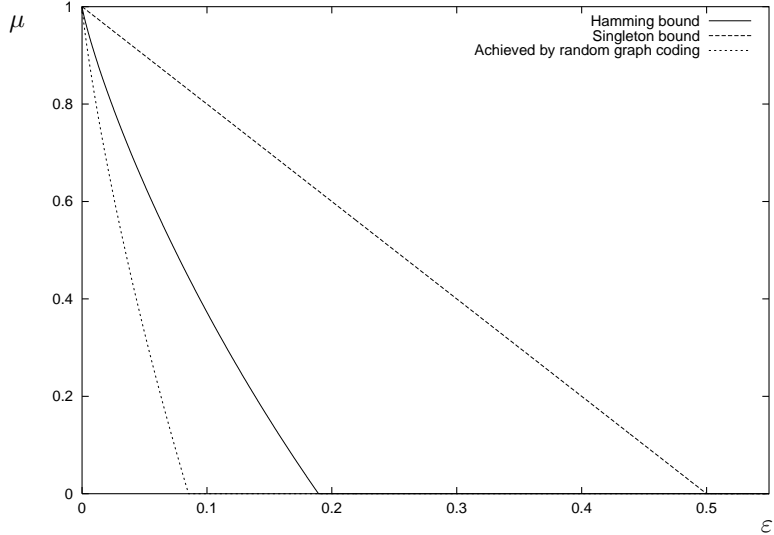}}
    \rput[r](13.5,0.5){$\epsilon$}
    \rput[tl](0.8,9){$\mu$}
  \end{pspicture}
    \caption{Singleton bound and Hamming bound together with the rate achieved by random graoh coding
      (for $d=2$). The allowed regions are below the respective curve.}
    \label{fig:bounds}
  \end{center}
\end{figure}

If we drop the nondegeneracy condition made above it is possible to find codes which break the Hamming
bound \cite{DiVSS}. In this case, however, we can consider the weaker \emph{singleton bound}, which has
to be respected by those \emph{degenerate codes} as well. It reads 
\begin{equation}
  1 - \frac{m}{n} \geq d \frac{f}{n}.
\end{equation}
We omit its proof here (see \cite{NC} Sect. 12.4 instead). Both bounds are plotted together with the rate
achieved by random graph coding in in Figure \ref{fig:bounds} (for $d=2$). 

%%%%%%%%%%%%%%%%%%%%%%%%%%%%%%%%%%%%%%%%%%%%%%%%%%%%%%%%%%%%%%%%%%%%%%%%%%%%%%%%%%%%%%%%%%%%%%%%%%%%
\section{Conclusions}
\label{sec:conclusions}
%%%%%%%%%%%%%%%%%%%%%%%%%%%%%%%%%%%%%%%%%%%%%%%%%%%%%%%%%%%%%%%%%%%%%%%%%%%%%%%%%%%%%%%%%%%%%%%%%%%%

We are now ready to combine our discussion of channel-capacity from
Section \ref{sec:channel-capacities} with the results about error
correction we have derived in the previous sections. Please note that most of the result presented here
can be found in \cite{Hamada01,MaUy01}, in some cases with better bounds.

%%%%%%%%%%%%%%%%%%%%%%%%%%%%%%%%%%%%%%%%%%%%%%%%%%
\subsection{Correcting small errors}
%%%%%%%%%%%%%%%%%%%%%%%%%%%%%%%%%%%%%%%%%%%%%%%%%%

We first look at the problem which motivated our study, namely
estimating the capacity of a channel $T\approx\Id$.

\begin{thm}\label{oneprime}
Let $d$ be a prime, and let $T$ be a channel on $d$-level systems.
Suppose that for some  $0 < \varepsilon < 1/2$,
\begin{equation}
  \cbnorm{\id-T}<2^{-H_2(\varepsilon)/\varepsilon}.
\end{equation}
Then
\begin{equation}\label{Qforprime}
  \QC(T)\geq (1-4\varepsilon)\log_2(d)-H_2(2\varepsilon)
\end{equation}
\end{thm}

\begin{proof}
For every $n$ set $f=\lfloor\varepsilon n\rfloor$, and
$m=\lfloor\mu n\rfloor-1$, where $\mu$ is, up to a $\log_2(d)$ factor, the right hand side of (\ref{Qforprime}), i.e. 
$\mu=1-4\varepsilon-\log_2(d)^{-1}H_2(2\varepsilon)$. This ensures
that the right hand side of (\ref{Pnotcorrect}) is strictly
negative, so there must be a code for $d$-level systems, with $m$
inputs and $n$ outputs, and correcting $f$ errors. To this code we
apply Proposition~\ref{Pinomi}, and insert the bound on
$\cbnorm{\id-T}$ into Equation (\ref{errorexpt}). Thus
$\Delta(T^{\otimes n},d^{\lfloor\mu n\rfloor-1})\to0$, even
exponentially. This means that any number $<\mu\log_2(d)$ is an
achievable rate. In other words, $\mu\log_2(d)$ is a lower bound
to the capacity.
\end{proof}

If $\epsilon > 0$ is small enough the quantity on the right hand side of Equation (\ref{Qforprime}) is strictly
positive (cf. the dotted graph in Figure \ref{fig:bounds}). Hence each channel which is sufficiently close
to the identity allows (asymptotically) perfect error correction. Beyond that we see immediately that
$\QC(T)$ is continous (in the cb-norm) at $T=\Id$: Since $\QC(T)$ is smaller than $\log_2(d)$ and $g(\epsilon)$ is
continuous in $\epsilon$ with $g(0)=\log_2(d)$ we find for each $\delta > 0$ an $\epsilon > 0$ exists, such that $\log_2(d)
- \QC(T) < \epsilon$ for all $T$ with $\|T-\Id\|_\cb < \epsilon/e$. In other words if $T$ is arbitrarily close to the
identity its capacity is arbitrarily close to $\log_2(d)$. In Corollary \ref{kor:1} below we will show the
significantly stronger statement that $Q$ is a lower semicontinuous function on the set of all channels.

%%%%%%%%%%%%%%%%%%%%%%%%%%%%%%%%%%%%%%%%%%%%%%%%%%
\subsection{Estimating capacity from finite coding solutions}
%%%%%%%%%%%%%%%%%%%%%%%%%%%%%%%%%%%%%%%%%%%%%%%%%%

A crucial consequence of the ability to correct small errors is
that we do not actually have to compute the limit defining the
capacity: if we have a pretty good coding scheme for a given
channel, i.e., one that gives us $ET^{\otimes n}D\approx\id_d$,
then we know the errors can actually be brought to zero, and the
capacity is close to the nominal rate of this scheme, namely
$\log_2(d)/n$.

\begin{thm} \label{prop:1}
Let $T$ be a channel, not necessarily between systems of the same
dimension. Let $k,p\in\Nl$ with $p$ a prime number, and 
suppose there are  channels $E$ and $D$ encoding and decoding a
$p$-level system through $k$ parallel uses of $T$, with error
$\Delta=\cbnorm{\id_p-ET^{\otimes k}D} < \frac{1}{2e}$. Then
\begin{equation}\label{Qforall}
  \QC(T)\geq \frac{\log_2(p)}n(1-4e\Delta)-\frac1nH_2(2e\Delta)\;.
\end{equation}
Moreover, $\QC(T)$ is the least upper bound on all expressions of
this form.
\end{thm}

\begin{proof}
We apply Proposition~\ref{oneprime} to the channel $\widetilde
T=ET^{\otimes n}D$. With the random coding method we thus find a
family of coding and decoding channels $\widetilde E$ and
$\widetilde D$ from $m'$ into $n'$ systems, of $p$ levels each, 
 such that
\begin{equation}
  \cbnorm{\id-\widetilde E
      \bigl(ET^{\otimes k}D\bigl)^{\otimes n'}
  \widetilde D}\to0.
\end{equation}
This can be reinterpreted as an encoding of $p^{m'}$-dimensional systems through $kn'$ uses of the channel
$T$ (rather than $\widetilde{T}$), which corresponds to a rate $(kn')^{-1}\log_2(p^{m'})=(\log_2 p/k)(m'/n')$.
We now argue exactly as in the proof of the previous proposition,
with $\varepsilon=e\Delta$, so that 
\begin{equation} \label{eq:9}
  \cbnorm{\id_p-ET^{\otimes k}D}=\varepsilon/e\leq 2^{H_2(\varepsilon)/\varepsilon}
\end{equation}
by equation (\ref{simplebound}). By random graph coding we can achieve the coding
ratio  $\mu\approx(m'/n') = 1-4\varepsilon-\log_2(p)^{-1}H_2(2\varepsilon)$, and have the errors $\Delta(\widetilde{T}^{\otimes n'},p^{m'})$ go to zero
exponentially. Since
\begin{equation} \label{eq:12}
  \Delta(T^{\otimes kn'},p^{m'}) \leq \Delta(\widetilde{T}^{\otimes n'},p^{m'}) \leq \cbnorm{\id-\widetilde E \bigl(ET^{\otimes k}D\bigl)^{\otimes n'}},
\end{equation}
we can apply Lemma \ref{lem:1} to the channel $T$ (where the sequence $n_\alpha$ is given by $n_\alpha = n\alpha$) and
find that the rate $\mu(\log_2 p/k)$ is achievable. This yields the estimate claimed in Equation
(\ref{Qforall}). 

To prove the second statement consider the function $x \to p(x)$ which associates to each real
number $x \geq 2$ the biggest prime $p(x)$ with $p(x) \leq x$. From known bounds on the length of gaps between
two consecutive primes \cite{Ingham37}\footnote{If $p_n$ denotes the $n^{\rm th}$ prime and $g(p_n) =
  p_{n+1} - p_n$ is the length of the gap between $p_n$ and $p_{n+1}$ it is shown in \cite{Ingham37} that
  $g(p)$ is bounded by ${\rm const} p^{5/8 + \epsilon}$.} it follows that $\lim_{x\to \infty} x/p(x) = 1$ holds, hence we
get $2^{kc}/p(2^{kc}) \leq 1 + \delta'$ for an arbitrary $\delta' > 0$, provided $n$ is large enough, but this implies
\begin{equation}
c - \frac{\log_2\bigl[p(2^{kc})\bigr]}{k}  < \frac{\log_2(1+\delta')}{k}. 
\end{equation}
Since we can choose an achievable rate $c$ arbitrarily close to the capacity $\QC(T)$ this shows that
there is for each $\delta > 0$ a prime $p$ and a positive integer $k$ such that $|\QC(T) - \log_2(p)/k| \leq
\delta$. In addition we can find a coding scheme $E$, $D$ for $T^{\otimes k}$ such that  Equation (\ref{eq:9})
holds, i.e. the right hand side of (\ref{Qforall}) can be arbitrarily close to  $\log_2(p)/k$, and this
completes the proof. 
\end{proof} 

This theorem allows us to derive very easily an important \emph{continuity property} of the quantum
capacity. It is well known that each function $F$ (on a topological space) which is given as the supremum
of a set of real-valued, continuous functions is \emph{lower semicontinuous}, i.e. the set
$F^{-1}\bigl((x,\infty]\bigr)$ is open for each $x \in \Bbb{R}$. Since the right hand side of Equation
(\ref{Qforall}) is continuous in $T$ and since $Q(T)$ is (according to Proposition \ref{prop:1}) the
supremum over such quantities, we get:  

\begin{kor}\label{kor:1}
$T\mapsto \QC(T)$ is lower semi-continuous in cb-norm.
\end{kor}

%%%%%%%%%%%%%%%%%%%%%%%%%%%%%%%%%%%%%%%%%%%%%%%%%%
\subsection{Error exponents}
%%%%%%%%%%%%%%%%%%%%%%%%%%%%%%%%%%%%%%%%%%%%%%%%%%

Another consequence of Theorem \ref{prop:1} concerns the rate with which the error $\Delta(T^{\otimes n},
2^{\lfloor cn\rfloor})$ decays in the limit $n \to \infty$. Theorem \ref{prop:1} says, roughly speaking that we can
achieve \emph{each rate} $c < \QC(T)$ by combining a coding scheme $E, D$ with subsequent random-graph
coding $\widetilde{E}, \widetilde{D}$. However, the error $ \Delta\bigl[(ET^{\otimes n}D)^{\otimes l},p^{k}\bigr]$
decays according to (\ref{errorexpt}) and Proposition \ref{prop:2} exponentially. A more precise analysis
of this idea leads to the following (cf. also the work Hamada \cite{Hamada01}):

\begin{prop} \label{prop:3}
  If $T$ is a channel with quantum capacity $\QC(T)$ and
  $c<\QC(T)$, then, for sufficiently large $n$ we have
  \begin{equation}
    \Delta(T^{\otimes n}, 2^{\lfloor cn\rfloor})\leq e^{- n \lambda(c)},
  \end{equation}
  with a positive constant $\lambda(c)$. 
\end{prop}

\begin{proof}
  We start as in Theorem \ref{prop:1} with the channel $\widetilde{T}=ET^{\otimes k}D$ and the quantity
  $\Delta=\cbnorm{\id_p-ET^{\otimes k}D}$. However instead of assuming that $\Delta = \epsilon/e$ holds, the full range $e\Delta \leq \epsilon
  \leq 1/2$ is allowed for the error rate $\epsilon$. Using the same arguments as in the proof of Theorem
  \ref{prop:1} we get an achievable rate
  \begin{equation} \label{eq:16}
    c(k,p,\epsilon) = \frac{\log_2(p)}{k} \left(1 - 4\epsilon - \frac{H_2(2\epsilon)}{\log_2(p)}\right)
  \end{equation}
  and an exponential bound on the coding error:
  \begin{equation} \label{eq:13}
    \Delta(T^{\otimes kn'},p^{m'})  \leq \cbnorm{\id-\widetilde E \bigl(ET^{\otimes k}D\bigl)^{\otimes n'}} \leq \left(2^{H_2(\epsilon)} \Delta^\epsilon\right)^{n'};
  \end{equation}
  cf. Equations (\ref{errorexpt}) and (\ref{eq:12}). 

  To calculate the exponential rate $\lambda(c)$ with which the coding error vanishes we have to consider the
  quantity 
  \begin{align} \label{eq:14}
    \lambda(c) &= \liminf_{n \to \infty} -\frac{1}{n} \ln \Delta(T^{\otimes n}, \lfloor2^{nc}\rfloor) \geq \lim_{n' \to \infty}\frac{-1}{kn'} n' \ln
    \left(2^{H_2(\epsilon)} \Delta^\epsilon\right) \\
    &\geq - \frac{\epsilon}{k} \left( \ln(\Delta) + \ln2\frac{H_2(\epsilon)}{\epsilon} \right)  = -\epsilon \Lambda(\Delta,\epsilon)/k \label{eq:15}
  \end{align}
  where we have inserted inequality (\ref{eq:13}). Now we we can apply Lemma \ref{lem:1} (with the
  sequence $n_\alpha = k\alpha$), which shows that $\lambda(c)$ is positive, if the right hand side of (\ref{eq:15})
  is. 

  What remains to show is that $\lambda(c) > 0$ holds for each $c < \QC(T)$. To this end we have to choose
  $k,p,\Delta$ and $\epsilon$ such that $c(k,p,\epsilon) = c$ and $\Lambda(\Delta,\epsilon) < 0$. Hence consider $\delta > 0$ such that
  $c + \delta < \QC(T)$ is an achievable rate. As in the proof of Theorem \ref{prop:1} we can choose
  $\log_2(p)/k$ such that $\log_2(p)/k > c + \delta$ holds while $\Delta$ is arbitrarily small. Hence there is an
  $\epsilon_0 > 0$ such that $c(k,p,\epsilon) = c$ implies $\epsilon > \epsilon_0$. The statement therefore follows from the fact
  that there is a $\Delta_0 > 0$ with $\Lambda(\Delta,\epsilon) > 0$ for all $0 < \Delta < \Delta_0$ and $\epsilon > \epsilon_0$. 
\end{proof} 

\begin{figure}[h]
  \begin{center}
    \begin{pspicture}(15,10)
    \rput(7.5,5){\includegraphics[scale=0.8]{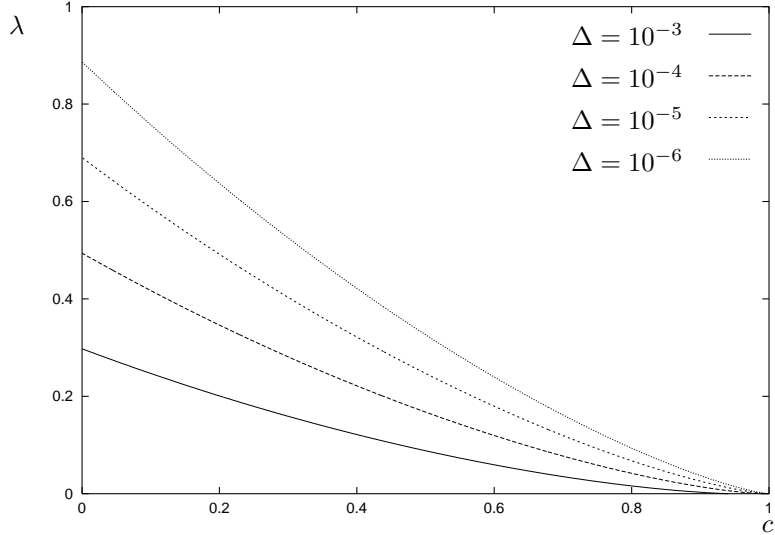}}
    \rput[r](13.5,0.5){$c$}
    \rput[tl](0.8,9){$\lambda$}
    \rput[r](12,8.7){$\Delta = 10^{-3}$}
    \rput[r](12,8){$\Delta = 10^{-4}$}
    \rput[r](12,7.3){$\Delta = 10^{-5}$}
    \rput[r](12,6.6){$\Delta = 10^{-6}$}
  \end{pspicture}
    \caption{Lower bounds on the error exponent $\lambda(c)$ plotted for $n=1, p=2$ and different values of $\Delta$.}
    \label{fig:errexp}
  \end{center}
\end{figure}

In addition to the statement of Proposition \ref{prop:3} we have just derived a lower bound on the error
exponent $\lambda(c)$. Since we can not express the error rate $\epsilon$ as a function of $k,p$ and $c$ we can not
specify this bound explicity. However we can plot it as a parametrized curve (using Equation
(\ref{eq:16}) and (\ref{eq:15}) with $\epsilon$ as the parameter) in the $(c,\lambda)$-space. In Figure
\ref{fig:errexp} this is done for $k=1$, $p=2$ and several values of $\Delta$.

%%%%%%%%%%%%%%%%%%%%%%%%%%%%%%%%%%%%%%%%%%%%%%%%%%
\subsection{Capacity with finite error allowed}
%%%%%%%%%%%%%%%%%%%%%%%%%%%%%%%%%%%%%%%%%%%%%%%%%%

We can also tolerate finite errors in encoding. Let
$Q_\varepsilon(T)$ denote the quantity defined exactly like the
capacity, but with the weaker requirement that $\Delta(T^{\otimes
n}, 2^{\lfloor cn\rfloor})\leq \varepsilon$ for large $n$. Obviously we have $\QC_\epsilon(T) \geq \QC(T)$ for each $\epsilon > 0$. Regarded as a
function of $\epsilon$ and $T$ this new quantity admits in addition the following continuity property in $\epsilon$.

\begin{prop} $\lim_{\varepsilon\to0}Q_\varepsilon(T)=\QC(T)$.
\end{prop} 

\begin{proof}
  By definition we can find for each $\epsilon',\delta > 0$ a tuple $n,p, E$ and $D$ such that
  \begin{equation}
    \cbnorm{\id_p-ET^{\otimes n}D}=\frac{\varepsilon'+\epsilon}{e}
  \end{equation}
  and $|\QC_\epsilon(T) - \log_2(p)/n| < \delta$ holds.  If $\epsilon+\epsilon'$ is small enough, however, we find as in Theorem
  \ref{prop:1} a random graph coding scheme such that
  \begin{equation}
    \QC(T)\geq \frac{\log_2(p)}n\bigl(1-4 (\epsilon+\epsilon')\bigr)-\frac{1}{n} H_2\bigr(2(\epsilon+\epsilon')\bigl) = g(\epsilon+\epsilon').
  \end{equation}
  Hence the statement follows from continuity of $g$ and the fact that $g(0) = \log_2(p)/n$ holds.
\end{proof}

For a classical channel $\Phi$ even more is known about the similar defined quantity $C_\epsilon(T)$: If $\epsilon > 0$ is
small enough we can not achieve bigger rates by allowing small errors, i.e. $C(T) = C_\epsilon(T)$. This is
called the ``strong converse of Shannon's noisy channel coding theorem'' \cite{Shannon48}. To check
whether a similar statement holds in the quantum case is one of the big open problem of the theory.  

\section*{Acknowledgements}

Funding by the European Union project EQUIP (contract IST-1999-11053)
and financial support from the DFG (Bonn) is greatfully acknowledged.

%\bibliographystyle{mk}
%\bibliography{qinf}

\end{document}